\pdfoutput=1
\documentclass[twocolumn,superscriptaddress,aps,preprintnumbers,amsmath,amssymb,prl,nofootinbib]{revtex4}

\usepackage{amsmath}
\usepackage{amssymb}
\usepackage{amsfonts}
\usepackage{graphicx}
\usepackage{xcolor}
\usepackage{xfrac}
\usepackage{comment}
\usepackage{pifont}
\usepackage{physics}
\usepackage{fourier}
\usepackage{hyperref}
\usepackage{bm}
\usepackage{enumitem}
\usepackage{xfrac}
\usepackage{tcolorbox}

\definecolor{rossoferrari}{HTML}{D9073D}
\definecolor{mediumblue}{HTML}{0000CD}
\definecolor{forestgreen}{HTML}{228B22}
\definecolor{desy_blue}{HTML}{009EE2}
\definecolor{desy_orange}{HTML}{FD8800}
\definecolor{light_pink}{rgb}{1,0.4,0.4}
\definecolor{light_blue}{rgb}{0.284602,0.317763,0.963947}
\hypersetup{setpagesize=false,bookmarksnumbered=true,bookmarksopen=true,%
colorlinks=true,linkcolor=light_blue,urlcolor=rossoferrari,citecolor=rossoferrari,linktocpage=false}

\begin{document}


\preprint{MS-TP-25-20}

\title{From new physics to a running power law and back again: Minimal refitting techniques for\\the reconstruction of the gravitational-wave background signal in pulsar timing array data}

\author{David Esmyol}
\email{d_esmy01@uni-muenster.de}
\affiliation{Institute for Theoretical Physics, University of M\"unster, 48149 M\"unster, Germany}

\author{Antonio J.\ Iovino}
\email{a.iovino@nyu.edu}
\affiliation{New York University, Abu Dhabi, PO Box 129188 Saadiyat Island, Abu Dhabi, UAE}
\affiliation{Institute for Theoretical Physics, University of M\"unster, 48149 M\"unster, Germany}

\author{Kai Schmitz}
\email{kai.schmitz@uni-muenster.de}
\affiliation{Institute for Theoretical Physics, University of M\"unster, 48149 M\"unster, Germany}
\affiliation{Kavli IPMU (WPI), UTIAS, The University of Tokyo, Kashiwa, Chiba 277-8583, Japan}



\begin{abstract}
Pulsar timing array (PTA) collaborations recently presented evidence for a gravitational-wave background (GWB) signal at nanohertz frequencies. In this paper, we introduce new refitting techniques for PTA data analysis that elevate related techniques in the literature to a more rigorous level and thus provide the basis for fast and accurate Bayesian inference and physically intuitive model comparisons. The key idea behind our approach is to construct maps $\Phi$ from GWB spectral models to a running-power-law (RPL) reference model, such that the pullback $\Phi^* P_{\rm RPL}$ of the RPL posterior density $P_{\rm RPL}$ induces a likelihood on the GWB model parameter space; in other words, we refit spectral models to the RPL posterior density. In order to construct $\Phi$, we introduce a matched-filtering approach in which $\Phi$ follows from a $\chi^2$ minimization that accounts for the frequency dependence of PTA sensitivity curves. We validate and illustrate our techniques by three concrete examples: GWs from stable cosmic strings, GWs from metastable strings, and scalar-induced GWs. 
\end{abstract}


\date{\today}
\maketitle


\noindent
\textbf{Introduction}\,---\,Einstein's theory of gravity predicts that a stochastic background of gravitational waves (GWs) washing through the Milky Way will imprint a correlated signature in the timing residuals of galactic millisecond pulsars. Pulsars timing arrays (PTAs)~\cite{Taylor:2021yjx,Postnov:2025lgd,Kelley:2025yud} are GW detectors of galactic dimensions that search for a signal of this type by monitoring the pulse arrival times from pulsars in our galactic neighborhood over many years or even decades. In summer 2023, PTA collaborations around the globe presented their latest data sets~\cite{NANOGrav:2023gor,EPTA:2023fyk,Reardon:2023gzh,Xu:2023wog,InternationalPulsarTimingArray:2023mzf} (see also Ref.~\cite{Miles:2024seg}), which showed for the first time evidence for the expected correlation pattern, i.e., the celebrated Hellings--Down curve~\cite{Hellings:1983fr,Romano:2023zhb}. It is expected that the significance of the observed correlation pattern will cross the $5\sigma$ threshold with future PTA data sets~\cite{NANOGrav:2020spf}, which means that we find ourselves at the brink of a detection of a GW background (GWB) signal at nanohertz frequencies. 

These exciting prospects have stimulated intensive research activities across different communities in recent years, aiming at elucidating the origin of the 2023 PTA signal. The most plausable explanation of the signal is of astrophysical nature and consists of a population of supermassive black-hole binaries~\cite{NANOGrav:2023hfp,Ellis:2023dgf,NANOGrav:2024nmo}. At the same time, the idea of a cosmological interpretation in terms of GWs from the early Universe~\cite{NANOGrav:2023hvm,EPTA:2023xxk,Figueroa:2023zhu,Ellis:2023oxs} has attracted much attention in the literature. While arguably more exotic than the astrophysical interpretation, a GW echo from the Big Bang would promise to open a new window onto the cosmology of our Universe at very early times and hence particle physics at very high energies~\cite{Maggiore:1999vm,Caprini:2018mtu}. Indeed, numerous exotic GWB models relying on new physics in the early Universe have been fitted to the 2023 PTA data, ranging from cosmic inflation~\cite{Vagnozzi:2023lwo,Jiang:2023gfe,Choudhury:2023kam} and scalar-induced GWs (SIGWs)~\cite{Franciolini:2023pbf,Cai:2023dls,Inomata:2023zup,Wang:2023ost,Firouzjahi:2023lzg,Liu:2023ymk,Balaji:2023ehk,Iovino:2024tyg,Franciolini:2023wjm} over cosmological phase transitions~\cite{Addazi:2023jvg,Bai:2023cqj,Han:2023olf,Megias:2023kiy,Ghosh:2023aum,Li:2023bxy,DiBari:2023upq,Gouttenoire:2023bqy,An:2023jxf} to cosmic defects such as cosmic strings~\cite{Ellis:2023tsl,Wang:2023len,Lazarides:2023ksx,Chowdhury:2023opo,Servant:2023mwt,Antusch:2023zjk,Ge:2023rce,Basilakos:2023xof} and domain walls~\cite{Kitajima:2023cek,Guo:2023hyp,Blasi:2023sej,Gouttenoire:2023ftk,Lu:2023mcz,Babichev:2023pbf}. 

Conclusively identifying the source of the PTA signal is a challenging task that will ultimately rely on the combination of different PTA observables, including an accurate measurement of the GWB frequency spectrum, GWB anisotropies across the sky~\cite{NANOGrav:2023tcn,Gardiner:2023zzr,Depta:2024ykq,Konstandin:2024fyo}, and possibly the detection of continuous GWs~\cite{NANOGrav:2023pdq,NANOGrav:2023wsz} and other multi-messenger signatures from individual binaries~\cite{DOrazio:2023rvl}. At present, however, in the absence of any anisotropies or continuous-wave signals, the premier PTA observable that is being used by particle physicists and astrophysicists alike for model discrimination and parameter inference is the GWB spectrum. 

Much of the particle physics literature on the 2023 PTA signal revolves around questions such as: Does a given GWB spectral model based on new physics beyond the Standard Model (BSM) yield a sufficient signal amplitude $A$? What is the spectral tilt $\gamma$ predicted by the model? Which parameter values yield the best agreement with the $A$ and $\gamma$ posteriors presented by this or that PTA collaboration? These questions offer a useful language to talk about the GWB spectra under consideration and help in developing a physical intuition for their properties. At the same time, it is clear that these questions do not represent much more than a vague mental picture; $A$ and $\gamma$ denote the parameters of one specific GWB spectral model, the constant power law (CPL), and do not exist in BSM models, such as phase-transition or cosmic-string models. Nearly all exotic models notably predict a GWB spectrum that deviates from a CPL spectrum, such that it is unclear how \textit{exactly} the predictions of exotic models should be mapped onto the CPL parameter space.

The aim of the present paper is to improve on this situation and put the mapping between BSM models and simple reference models auch as the CPL model on a more rigorous footing. Specifically, we shall introduce a new approach based on the following three ingredients: \textbf{\ding{117}}~Instead of the CPL, we will work with a running power law (RPL) as our physics-agnostic reference model. The RPL allows for a logarithmic running of the spectral index and hence better reflects the rough shape of the GWB spectrum that one finds in many BSM models. \textbf{\ding{117}}~We construct the map $\Phi$ between a given BSM model and the RPL in terms of a $\chi^2$ function that accounts of the frequency dependence of PTA sensitivity curves. In this way, we ensure that our map $\Phi$ is based on global information, i.e., the GWB spectrum across the whole PTA frequency band, and not just on local information at an arbitrarily chosen pivot frequency. \textbf{\ding{117}}~We go from BSM models to the RPL, and back again\,---\,which is to say that we do not content ourselves with a simple projection onto the RPL parameter space. Instead, we construct the pullback $\Phi^*P_{\rm RPL}$ of the RPL posterior density $P_{\rm RPL}$, which induces a likelihood on the BSM parameter space and thus provides the basis for simple spectral refits. We will validate and illustrate our method by explicitly refitting three GWB spectral models to the RPL posterior density: stable cosmic strings~\cite{Sousa:2024ytl,Schmitz:2024gds}, metastable strings~\cite{Buchmuller:2021mbb,Buchmuller:2023aus}, and SIGWs~\cite{Domenech:2021ztg}.

Finally, before we present our approach in detail, we caution that, even with the improvements listed above, spectral refits are and remain an approximate method. Our approach of refitting the RPL posterior density is similar in spirit to refits of the amplitude posterior densities (``violins'') in the free-spectral model, which are, e.g., implemented in the software package \texttt{ceffyl}~\cite{Lamb:2023jls}. Any more rigorous analysis should directly resort to Bayesian MCMC fits to PTA timing residuals, e.g., using tools like \texttt{enterprise}~\cite{2019ascl.soft12015E}, \texttt{enterprise\_extensions}~\cite{enterprise}, and \texttt{PTArcade}~\cite{Mitridate:2023oar}. As part of our numerical analysis, we will quantify the agreement between our approximate results and full MCMC runs.


\smallskip\noindent
\textbf{\ding{117}~Running power law (RPL)}\,---\,The spectral composition of the GWB can be quantified in terms of several different quantities; in this paper, we shall work with the GWB energy density power spectrum on a logarithmic frequency axis in units of the critical energy density of the Universe~\cite{Maggiore:2007ulw,Maggiore:2018sht},
\begin{equation}
\Omega_{\rm GW}\left(f\right) = \frac{1}{\rho_{\rm crit}}\,\frac{d\rho_{\rm GW}}{d \ln f} \,.
\end{equation}
In the RPL model, $\Omega_{\rm GW}$ is modeled as a power law with a logarithmically running spectral index, i.e., as a parabola-shaped spectrum if plotted on log--log axes,
\begin{equation}
\Omega_{\rm RPL}\left(f\right) = \frac{2\pi^2}{3H_0^2}\,A^2 f_{\rm ref}^2\left(\frac{f}{f_{\rm ref}}\right)^{5-\left[\gamma + (\beta/2)\,\ln\left(f/f_{\rm ref}\right) \right]} \,.
\end{equation}
Here, $H_0 = 100\,h\,\textrm{km}/\textrm{s}/\textrm{Mpc}$ is the Hubble constant, where we set $h= 0.674$, following Ref.~\cite{ngcurve}; $f_{\rm ref}$ is an arbitrary reference scale, which we set to $f_{\rm ref} = \sfrac{1}{10\,\textrm{yr}}$, following Ref.~\cite{Agazie:2024kdi}; and $A$, $\gamma$, and $\beta$ are the three parameters of the RPL model: $A$ and $\gamma$ control the amplitude and the spectral index of $\Omega_{\rm GW}$ at $f = f_{\rm ref}$, respectively, while $\beta$ controls the constant logarithmic running of the spectral index at all frequencies. Correspondingly, setting $\beta = 0$ returns the CPL model,
\begin{equation}
\Omega_{\rm CPL}\left(f\right) = \frac{2\pi^2}{3H_0^2}\,A^2 f_{\rm ref}^2\left(\frac{f}{f_{\rm ref}}\right)^{5-\gamma} \,.
\end{equation}

For our spectral refits of the RPL model, we require the posterior density $P_{\rm RPL}$ on the three-dimensional RPL model parameter space after fitting the model to PTA timing residuals. A Bayesian fit of this type has been carried out in Ref.~\cite{Agazie:2024kdi}, where the RPL model was fitted to the NANOGrav 15-year (NG15) data set. In the following, we will work with the Markov Chain Monte Carlo (MCMC) data produced by this analysis and reconstruct $P_{\rm RPL}$ from these data in terms of a kernel density estimation (KDE). The analysis in Ref.~\cite{Agazie:2024kdi} assumed uniform priors for $\log_{10}A$, $\gamma$, and $\beta$, which is the set of parameters that we will refer to as the RPL model parameters in the following, $\bm{\theta}_{\rm RPL} = \left\{\log_{10}A,\gamma,\beta\right\}$. 

We emphasize that the KDE approximation of $P_{\rm RPL}$ is at the heart of our refitting analysis. As we will demonstrate, the spectral information stored in it turns out to suffice for an accurate parameter reconstruction across a whole range of exotic GWB models. Our spectral refits of the RPL model thus require only one full Bayesian MCMC fit to PTA timing residuals\,---\,the one in Ref.~\cite{Agazie:2024kdi}\,---\,based on which one is able to construct $P_{\rm RPL}$ \textit{once and for all} as the universal starting point for any RPL refit that one may be interested in.


\begin{figure}
\includegraphics[width=0.45\textwidth]{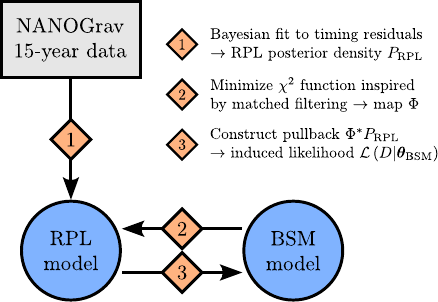}
\caption{Schematic illustration of our minimal refitting approach.}
\label{fig:cartoon}
\end{figure}


\begin{figure*}
\includegraphics[width=0.95\textwidth]{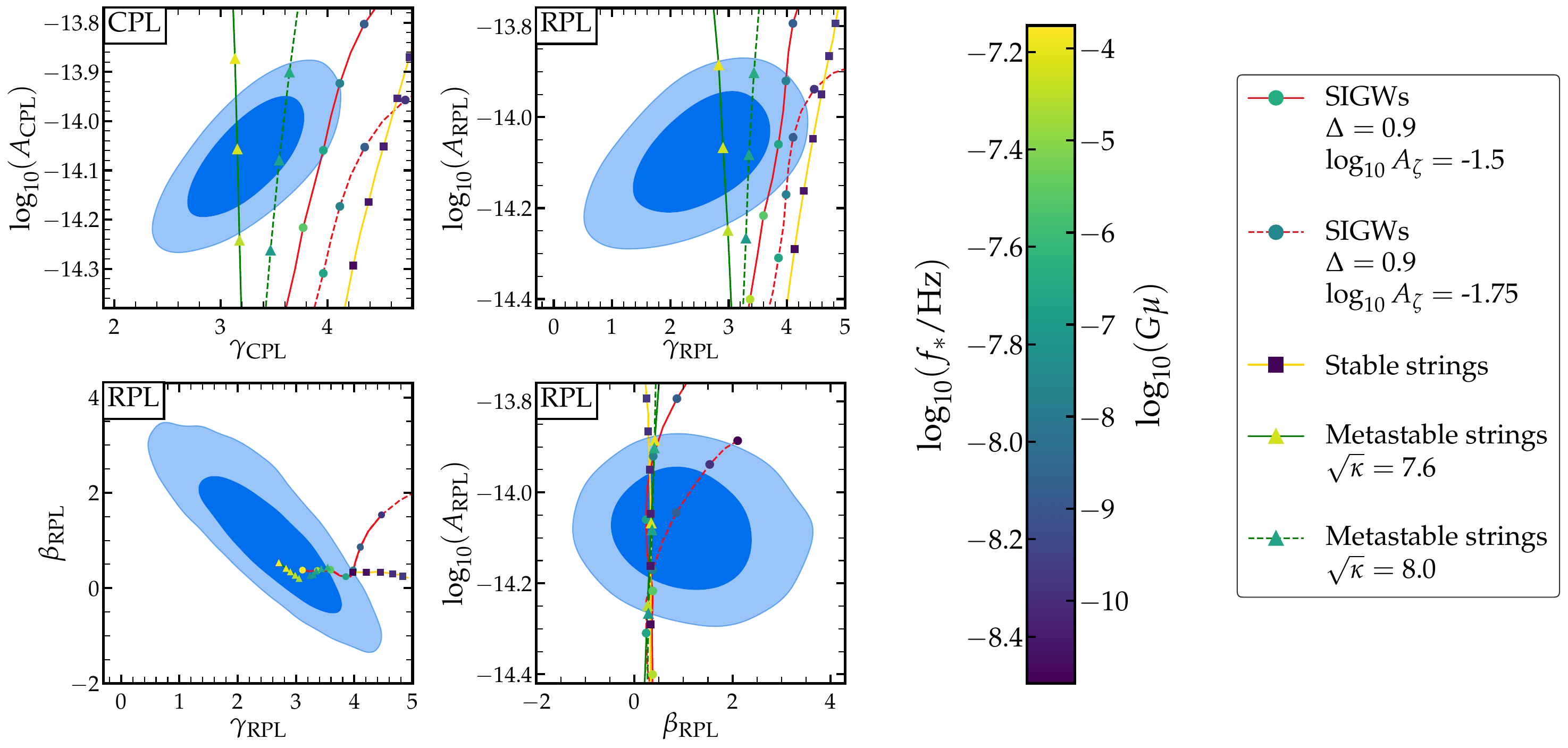}
\caption{Marginalized two-dimensional posterior densities for the parameters of the CPL and RPL models after Bayesian MCMC fits to the NG15 timing residuals based on the analysis in Ref.~\cite{Agazie:2024kdi}; $68\,\%$ and $95\,\%$ credible regions in dark and light blue, respectively. On top, we show projections of select exotic GWB spectra onto the parameter spaces of these two models constructed in terms of our map $\Phi$ in Eq.~\eqref{eq:Phi}. From the color scale, one can read off both $\log_{10}\left(f_*/\textrm{Hz}\right)$ in the SIGWs model as well as $\log_{10}\left(G\mu\right)$ for stable and metastable strings.}
\label{fig:parameters}
\end{figure*}


\smallskip\noindent
\textbf{\ding{117}~Matched filtering}\,---\,Next, we need to define the map from BSM models to the RPL baseline model. To do so, we start from the signal-noise-ratio (SNR) in Refs.~\cite{Hazboun:2019vhv,Schmitz:2020syl},
\begin{equation}
\label{eq:SNR}
\textrm{SNR}^2 = 2T \int_{f_{\rm min}}^{f_{\rm max}} \left(\frac{\Omega_{\rm GW}\left(f\right)}{\Omega_{\rm sens}\left(f\right)}\right)^2  \textrm{d}f \,,
\end{equation}
where $T$ is the total observing time, $\left[f_{\rm min},f_{\rm max}\right]$ the PTA frequency band, $\Omega_{\rm GW}$ the GW energy density power spectrum of the stochastic GWB signal of interest, and $\Omega_{\rm sens}$ the effective strain noise power spectrum of the PTA (i.e., the PTA sensitivity curve). The SNR in Eq.~\eqref{eq:SNR} represents the \textit{optimal expected} SNR that can be achieved in a PTA measurement: it is based on an optimal choice of filter function (i.e., a matched filter) in the construction of the cross-correlation statistic, and it corresponds to an expectation value (i.e., the average over all realizations of a stochastic GWB).

Assuming Gaussian detector noise, the matched-filtering procedure leading to Eq.~\eqref{eq:SNR} can be reformulated as the maximization of a Gaussian likelihood $\mathcal{L}$ (see Chapter~7 in Ref.~\cite{Maggiore:2007ulw}), or equivalently, as the minimization of a $\chi^2$ function, $\Delta\chi^2 = -2\,\ln\left(\mathcal{L}/\mathcal{L}_{\rm max}\right)$~\cite{Kuroyanagi:2018csn}. In our analysis, we shall work with $\Delta\chi^2$, which takes two types of GW spectra as input: the true (or fiducial) spectrum contained in the data and a template spectrum used in the reconstruction of this signal. In order to construct our BSM--RPL map, we shall identify the former with the GW spectrum predicted by the BSM model, $\Omega_{\rm BSM}$, and the latter with the GW spectrum predicted by the RPL model, $\Omega_{\rm RPL}$, which allows us to write
\begin{equation}
\label{eq:Deltachi2}
\Delta\chi^2 = 2T \int_{f_{\rm min}}^{f_{\rm max}} \left(\frac{\Omega_{\rm BSM}\left(f;\bm{\theta}_{\rm BSM}\right)-\Omega_{\rm RPL}\left(f;\bm{\theta}_{\rm RPL}\right)}{\Omega_{\rm sens}\left(f\right)}\right)^2 \textrm{d}f \,.
\end{equation}
Similar $\Delta\chi^2$ functions have been considered in the literature before: in Ref.~\cite{Kuroyanagi:2018csn}, our primary source of inspiration, where it is used for a Fisher forecast, but also, e.g., in Refs.~\cite{Caldwell:2018giq,DEramo:2019tit}. Furthermore, we note that the comparison between Eqs.~\eqref{eq:SNR} and \eqref{eq:Deltachi2} suggests a straightforward interpretation of our $\Delta\chi^2$ function: it can be thought of as an SNR value for the differential spectrum $\Delta \Omega_{\rm GW} = \Omega_{\rm BSM}-\Omega_{\rm RPL}$ given an experimental sensitivity quantified in terms of the sensitivity curve $\Omega_{\rm sens}$.


This interpretation makes it clear that the $\Delta\chi^2$ function in Eq.~\eqref{eq:Deltachi2} provides us with exactly what we need: in order to map the BSM model onto the RPL model, we want to find the RPL parameters $\bm{\theta}_{\rm RPL}$ that minimize $\left|\Delta \Omega_{\rm GW}\right|$ for given BSM parameters $\bm{\theta}_{\rm BSM}$, especially at the GW frequencies $f$ where the experimental sensitivity is the best. The $\Delta\chi^2$ function in Eq.~\eqref{eq:Deltachi2} allows us to formulate this optimization problem in the form of a $\chi^2$-minimization problem, i.e., we are interested in the RPL parameters $\bm{\theta}_{\rm RPL}^{\rm min}$ that minimize the SNR value for the differential spectrum $\Delta \Omega_{\rm GW}$,
\begin{equation}
\Delta\chi^2\left(\bm{\theta}_{\rm BSM},\bm{\theta}_{\rm RPL}^{\rm min}\right) = \min_{\bm{\theta}_{\rm RPL}} \,\Delta\chi^2\left(\bm{\theta}_{\rm BSM},\bm{\theta}_{\rm RPL}\right) \,,
\end{equation}
which allows us to define our BSM--RPL map $\Phi$ as the map from the parameters $\bm{\theta}_{\rm BSM}$ to the best-fit parameters $\bm{\theta}_{\rm RPL}^{\rm min}$,
\begin{equation}
\label{eq:Phi}
\Phi \quad:\quad \bm{\theta}_{\rm BSM} \quad\mapsto\quad \Phi\left(\bm{\theta}_{\rm BSM}\right) = \bm{\theta}_{\rm RPL}^{\rm min} \,.
\end{equation}

In our numerical analysis, we will evaluate $\Delta\chi^2$ in Eq.~\eqref{eq:Deltachi2} working with the NG15 sensitivity curve in Ref.~\cite{ngcurve}. Correspondingly, we will set $T = 16.03\,\textrm{yr}$ (i.e., the total observing span of the NG15 data set) and $f_{\rm min} = \sfrac{1}{T}$ and $f_{\rm max} = \sfrac{30}{T}$, respectively (i.e., the frequency range considered in Ref.~\cite{Agazie:2024kdi}).


\begin{figure*}
\includegraphics[width=0.92\textwidth]{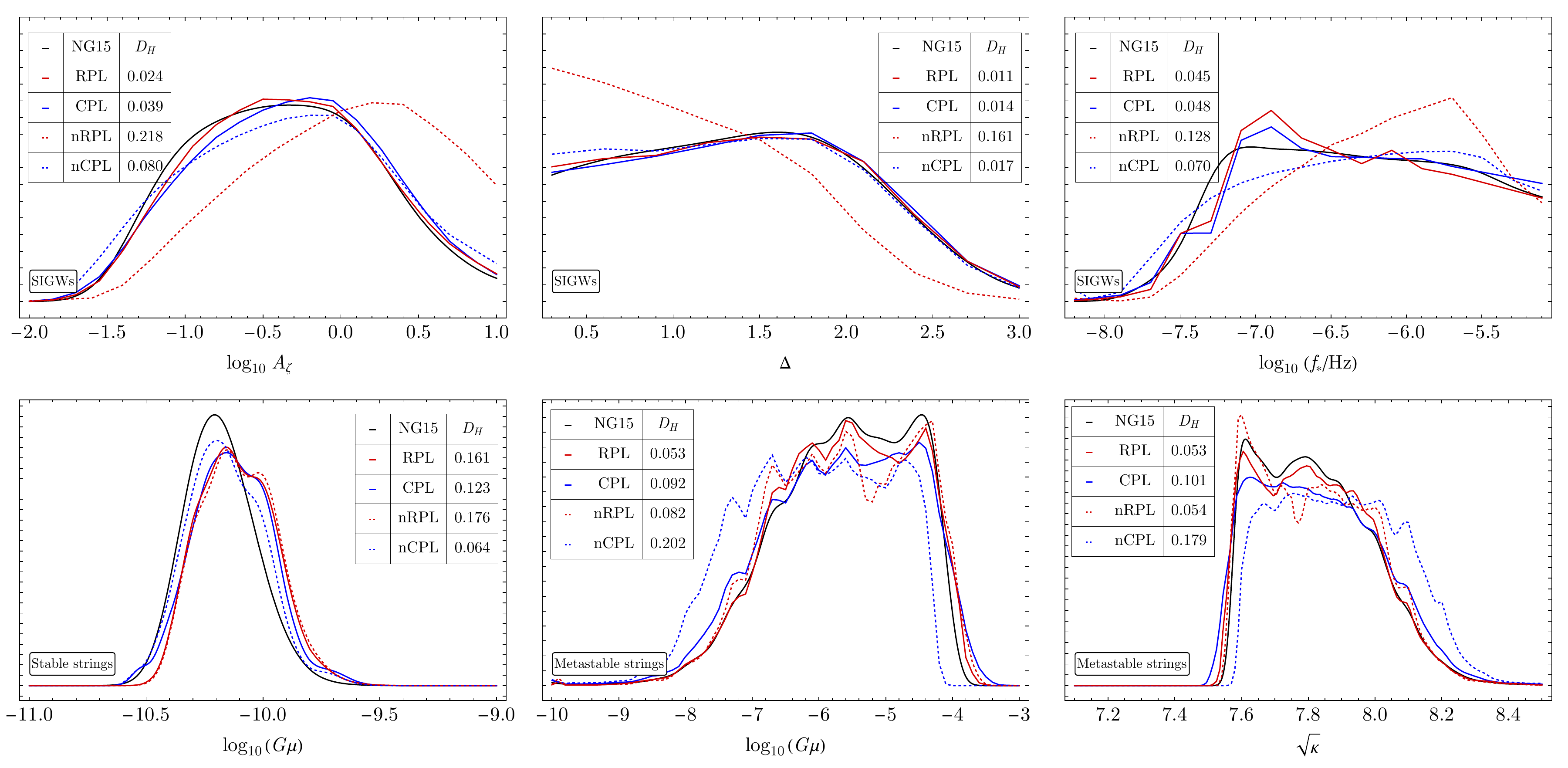}
\caption{Marginalized one-dimensional posterior densities for the parameters of three exotic GWB spectral models (SIGWs, stable strings, metastable strings) obtained in five different ways: \textbf{(NG15)} Bayesian MCMC fits to the NG15 timing residuals~\cite{NANOGrav:2023hvm}; \textbf{(RPL, nRPL)} refits to the RPL posterior density, based on the map $\Phi$ in Eq.~\eqref{eq:Phi} and the naive map in Eq.~\eqref{eq:nRPL}; \textbf{(CPL, nCPL)} analogous CPL refits, see Eqs.~\eqref{eq:Phi} and \eqref{eq:nCPL}. The tables list the Hellinger distances $D_H$ between the respective approximate posterior densities and the full NG15 MCMC results.}
\label{fig:posteriors}
\end{figure*}


\smallskip\noindent
\textbf{\ding{117}~Induced likelihood}\,---\,The third and last novel ingredient of our proposed minimal refitting approach consists in the construction of an induced likelihood on the BSM parameter space. In the first step of this construction, we treat the RPL parameters $\bm{\theta}_{\rm RPL}$ as latent parameters, which allows us to write the likelihood of the BSM model as follows,
\begin{equation}
\mathcal{L}\left(D|\bm{\theta}_{\rm BSM}\right) = \int \mathcal{L}\left(D|\bm{\theta}_{\rm RPL}\right) p\left(\bm{\theta}_{\rm RPL}|\bm{\theta}_{\rm BSM}\right) \, \textrm{d}\bm{\theta}_{\rm RPL} \,,
\end{equation}
where $D$ denotes the PTA data of interest, i.e., the NG15 data in our case, and the conditional probability $p\left(\bm{\theta}_{\rm RPL}|\bm{\theta}_{\rm BSM}\right)$ is a precursor of our BSM--RPL map. In the next step, we use Bayes' theorem to express the RPL posterior density as
\begin{equation}
P\left(\bm{\theta}_{\rm RPL}|D\right) = \frac{\mathcal{L}\left(D|\bm{\theta}_{\rm RPL}\right)\pi\left(\bm{\theta}_{\rm RPL}\right)}{Z_{\rm RPL}} \,,
\end{equation}
where $P_{\rm RPL} \equiv P\left(\bm{\theta}_{\rm RPL}|D\right)$, $\pi\left(\bm{\theta}_{\rm RPL}\right)$ is the prior density for the RPL parameters, and $Z_{\rm RPL}$ is the evidence of the RPL model,
\begin{equation}
Z_{\rm RPL} = \int \mathcal{L}\left(D|\bm{\theta}_{\rm RPL}\right)\pi\left(\bm{\theta}_{\rm RPL}\right) \, \textrm{d}\bm{\theta}_{\rm RPL}\,.
\end{equation}
Given the flat prior choice $\pi_{\rm RPL} \equiv \pi\left(\bm{\theta}_{\rm RPL}\right)$ in Ref.~\cite{Agazie:2024kdi}, the likelihood of the BSM model can be brought into the form
\begin{equation}
\mathcal{L}\left(D|\bm{\theta}_{\rm BSM}\right) = \frac{Z_{\rm RPL}}{\pi_{\rm RPL}}\int P\left(\bm{\theta}_{\rm RPL}|D\right) p\left(\bm{\theta}_{\rm RPL}|\bm{\theta}_{\rm BSM}\right) \, \textrm{d}\bm{\theta}_{\rm RPL} \,.
\end{equation}
Finally, we identify $p\left(\bm{\theta}_{\rm RPL}|\bm{\theta}_{\rm BSM}\right)$ as a Dirac delta function,
\begin{equation}
p\left(\bm{\theta}_{\rm RPL}|\bm{\theta}_{\rm BSM}\right) = \delta^{(3)}\left(\bm{\theta}_{\rm RPL}-\Phi\left(\bm{\theta}_{\rm BSM}\right)\right) \,,
\end{equation}
which leads to our final result for the likelihood $\mathcal{L}\left(D|\bm{\theta}_{\rm BSM}\right)$,
\begin{equation}
\mathcal{L}\left(D|\bm{\theta}_{\rm BSM}\right) = \frac{Z_{\rm RPL}}{\pi_{\rm RPL}}\,\left[\left(P_{\rm RPL}\circ \Phi\right)\left(\bm{\theta}_{\rm BSM}\right)\right] \,.
\end{equation}
Up to the constant prefactor $Z_{\rm RPL}/\pi_{\rm RPL}$, which is irrelevant for our purposes in this paper, the induced likelihood $\mathcal{L}\left(D|\bm{\theta}_{\rm BSM}\right)$ thus turns out to be nothing but the pullback of the RPL posterior density onto the BSM parameter space,
\begin{equation}
\label{eq:LBSM}
\mathcal{L}\left(D|\bm{\theta}_{\rm BSM}\right) \propto \left(P_{\rm RPL}\circ \Phi\right)\left(\bm{\theta}_{\rm BSM}\right) = \left(\Phi^*P_{\rm RPL}\right)\left(\bm{\theta}_{\rm BSM}\right) \,.
\end{equation}


This likelihood provides the starting point for parameter inference in the BSM model. Using again Bayes' theorem,
\begin{equation}
\label{eq:PBSM}
P\left(\bm{\theta}_{\rm BSM}|D\right) = \frac{\left[\left(P_{\rm RPL}\circ \Phi\right)\left(\bm{\theta}_{\rm BSM}\right)\right]\pi\left(\bm{\theta}_{\rm BSM}\right)}{\int \left[\left(P_{\rm RPL}\circ \Phi\right)\left(\bm{\theta}_{\rm BSM}\right)\right]\pi\left(\bm{\theta}_{\rm BSM}\right)\,\textrm{d}\bm{\theta}_{\rm BSM}} \,.
\end{equation}
For BSM models with a low-dimensional parameter space, one may numerically evaluate this relation by brute force. In fact, this is what we will do below. In more complicated cases, however, Eq.~\eqref{eq:PBSM} also lends itself to an evaluation in terms of more advanced techniques such as nested sampling. Furthermore, we note that Eq.~\eqref{eq:PBSM} also applies to BSM models with more than three parameters, i.e., more parameters than the RPL model, in which case the map $\Phi$ may no longer be invertible. Indeed, our approach does not require one to construct the inverse map $\Phi^{-1}$, which avoids issues with ``parameter degeneracies'' that other approaches in the literature sometimes run into; see, e.g., Ref.~\cite{Caprini:2024hue}.


\smallskip\noindent
\textbf{RPL refits}\,---\,The expression for the BSM posterior density in Eq.~\eqref{eq:PBSM} encapsulates the main message of our paper: the map $\Phi$, which follows from the minimization of a $\chi^2$ function based on matched filtering, allows us to project the predictions of BSM models onto the RPL parameter space; and the pullback $\Phi^* P_{\rm RPL}$ enables us to go back again from the RPL to the BSM model and construct an induced likelihood on the BSM parameter space; see Fig.~\ref{fig:cartoon} for an illustration.

We shall now illustrate the utility of our minimal refitting approach and apply it to three BSM models that were investigated in the search for BSM signals in the NG15 data~\cite{NANOGrav:2023hvm}:
\begin{enumerate}
\item \textit{Stable strings} (\textsc{stable-n} in Ref.~\cite{NANOGrav:2023hvm}):  One model parameter ($G\mu$, the string tension in units of Newton's constant), GW emission from cusps and kinks~\cite{Blanco-Pillado:2011egf,Blanco-Pillado:2013qja,Blanco-Pillado:2017oxo}.
\item \textit{Metastable strings} (\textsc{meta-ls} in Ref.~\cite{NANOGrav:2023hvm}): Two model parameters ($G\mu$ and $\sqrt{\kappa}$, the decay parameter controlling the lifetime of the string network), GW emission from cusps on loops and from segments~\cite{Buchmuller:2021mbb}.
\item \textit{SIGWs} (\textsc{sigw-gauss} in Ref.~\cite{NANOGrav:2023hvm}): Three model parameters ($A_\zeta$, $\Delta$, and $f_*$, which control the amplitude, width, and peak frequency of the primordial curvature spectrum $P_\zeta$, respectively), $P_\zeta$ is assumed to have a log-normal shape~\cite{Pi:2020otn}, $\Omega_{\rm GW}$ based on Refs.~\cite{Kohri:2018awv,Espinosa:2018eve}. 
\end{enumerate}

More details on these models and the computation of the respective GWB spectra can be found in Ref.~\cite{NANOGrav:2023hvm} and references therein. Ref.~\cite{NANOGrav:2023hvm} presents, moreover, the results of MCMC fits of all three models to the NG15 timing residuals, including the marginalized one-dimensional posterior densities for all parameters listed above. In the following, we shall reconstruct these posterior densities in four different ways and compare them to the results in Ref.~\cite{NANOGrav:2023hvm} \textbf{(NG15)}:
\begin{description}
\item[(RPL)] Refits to the RPL posterior density: explicit evaluation of Eq.~\eqref{eq:PBSM}, using the map $\Phi$ in Eq.~\eqref{eq:Phi}. These refits are the main results of our analysis in this paper.
\item[(CPL)] Refits to the CPL posterior density: same strategy as for our RPL refits, but with RPL $\rightarrow$ CPL in all steps of the analysis. With these refits, we intend to assess the relevance of nonzero $\beta$ in the baseline model.
\item[(nRPL)] Naive refits to the RPL posterior density: explicit evaluation of Eq.~\eqref{eq:PBSM}, but instead of $\Phi$, we use the BSM--RPL map that results from solving, for given $\bm{\theta}_{\rm BSM}$, the following constraints for $\bm{\theta}_{\rm RPL}$ at $f_{\rm pivot} = \sfrac{1}{10\,\textrm{yr}}$,
\begin{equation}
\label{eq:nRPL}
\left.\frac{d^n\ln\Omega_{\rm BSM}}{d\left(\ln f\right)^n}\right|_{f_{\rm pivot}} = \left.\frac{d^n\ln\Omega_{\rm RPL}}{d\left(\ln f\right)^n}\right|_{f_{\rm pivot}} \,, \quad n = 0,1,2 \,.
\end{equation}
\item[(nCPL)] Naive refits to the CPL posterior density: same strategy as for our naive RPL refits, but with RPL $\rightarrow$ CPL in all steps of the analysis. For given $\bm{\theta}_{\rm BSM}$, we solve the following constraints for $\bm{\theta}_{\rm CPL}$ at $f_{\rm pivot} = \sfrac{1}{10\,\textrm{yr}}$,
\begin{equation}
\label{eq:nCPL}
\left.\frac{d^n\ln\Omega_{\rm BSM}}{d\left(\ln f\right)^n}\right|_{f_{\rm pivot}} = \left.\frac{d^n\ln\Omega_{\rm CPL}}{d\left(\ln f\right)^n}\right|_{f_{\rm pivot}} \,, \qquad n = 0,1 \,.
\end{equation}
\end{description}

With the last two approaches, we intend to assess the relevance of the $\chi^2$-minimization procedure in the construction of our map $\Phi$. Note that the naive maps defined by Eqs.~\eqref{eq:nRPL} and \eqref{eq:nCPL} depend on the arbitrary choice of $f_{\rm pivot} \in \left[f_{\rm min},f_{\rm max}\right]$, whereas our map $\Phi$ is based on global information across the whole frequency band from $f_{\rm min}$ to $f_{\rm max}$. In our analysis, we set $f_{\rm pivot} = f_{\rm ref} = \sfrac{1}{10\,\textrm{yr}}$. This choice is convenient, but not necessary. It is still straightforward to solve the constraints in Eqs.~\eqref{eq:nRPL} and \eqref{eq:nCPL}, even if  $f_{\rm pivot} \neq f_{\rm ref}$.

In passing, we mention that the nCPL refits correspond to the standard procedure in the literature on the cosmic microwave background (CMB) and the predictions of individual models of cosmic inflation for the CMB observables $A_s$, $n_s$, and $r$ (i.e., the amplitude and tilt of the primordial scalar spectrum as well as the tensor-to-scalar ratio at some scale $k_{\rm pivot}$)~\cite{Planck:2018jri}. In CMB analyses of this type, the nCPL approach is justified by the slow-roll approximation, which can be systematically improved order by order in the slow-roll parameters. For instance, at next-to-leading order, one has to include the running $\alpha_s$ of the spectral index $n_s$, which then results in analyses on par with our nRPL refits. In our case, by contrast, there is no equivalent approximation. In other words, there are no small ``slow-roll'' parameters that would control the deviations between BSM spectra $\Omega_{\rm BSM}$ and the reference $\Omega_{\rm RPL}$ (or $\Omega_{\rm CPL}$) spectra, which motivates us to go beyond the naive maps in Eqs.~\eqref{eq:nRPL} and \eqref{eq:nCPL}.


\medskip\noindent
\textbf{Results}\,---\,We present the results of our numerical analysis in Figs.~\ref{fig:parameters} and \ref{fig:posteriors}. In Fig.~\ref{fig:parameters}, we show the projection of select exotic GWB spectra onto the CPL and RPL parameter spaces constructed in terms of our map $\Phi$ in Eq.~\eqref{eq:Phi}. The plots of the $\gamma$--$\log_{10}A$ plane can notably be regarded as the equivalents of the standard $n_s$--$r$ plot in the CMB literature. An important observation in view of our two plots of the $\gamma$--$\log_{10}A$ plane is that they differ not only in terms of the blue posterior ellipses, the model predictions indicated by the colored lines and markers also slightly change. The reason for this is clear: unlike in the nCPL and nRPL approaches, we do not Taylor-expand the $\Omega_{\rm BSM}$ spectra around some $f_{\rm pivot}$. This can lead to situations where, e.g., the BSM--RPL map results in a slightly different index at $f_{\rm ref}$ than the BSM--CPL map.

The outcome of our four different refit analyses (RPL, CPL, nRPL, nCPL) are shown in Fig.~\ref{fig:posteriors}, where we compare all approximate parameter posterior densities to the MCMC results from Ref.~\cite{NANOGrav:2023hvm} in terms of the Hellinger distance $D_H$. For metastable strings and SIGWs, we find that the RPL refits always yield the best results, i.e., the smallest Hellinger distances to the NG15 posteriors. Similarly, for both RPL and CPL, the naive refits always perform worse in these models than their more sophisticated counterparts, i.e., $D_H\left(\textrm{nRPL}\right) > D_H\left(\textrm{RPL}\right)$ and $D_H\left(\textrm{nCPL}\right) > D_H\left(\textrm{CPL}\right)$. These observations confirm our expectations and demonstrate that the RPL refits introduced in this paper offer a more accurate refitting technique than the simpler nRPL, CPL, and nCPL techniques. At the same time, our refits of the stable-strings model to the RPL and CPL posterior densities results in a different hierarchy, $D_H\left(\textrm{nCPL}\right) < D_H\left(\textrm{CPL}\right) < D_H\left(\textrm{RPL}\right) < D_H\left(\textrm{nRPL}\right)$, which we attribute to the poor quality of the stable-strings fit of the NG15 data in general; see Fig.~\ref{fig:parameters}, where the predictions of this model do not even intersect the $95\,\%$ credible regions for $\gamma$ and $\log_{10}A$. 

Finally, we comment on two technical numerical aspects: First, from the posterior densities for the SIGWs model, one can clearly infer the finite resolution of our scan in the $A_\zeta$--$\Delta$--$f_*$ parameter space. Nested sampling would provide an alternative approach to evaluating Eq.~\eqref{eq:PBSM}, especially for models with three or more parameters. Second, all posterior densities for metastable strings feature numerical fluctuations that ultimately result from the coarse resolution of the $G\mu$--$\sqrt{\kappa}$ parameter grid used in the analysis in Ref.~\cite{NANOGrav:2023hvm}. In future work, it would be interesting to recompute all posterior densities for metastable strings, including full MCMC results, with a finer resolution in parameter space. On the other hand, for our purposes in this paper, the numerical fluctuations in the posterior densities for metastable strings actually offer an unintended advantage, as they allow us to demonstrate that even very nontrivial posterior shapes can be reproduced by our minimal refitting techniques.


\medskip\noindent
\textbf{Conclusions}\,---\,We introduced a new minimal refitting technique, based on the RPL baseline model and a $\chi^2$-minimization procedure inspired by matched filtering, that can be used for model comparison (see Fig.~\ref{fig:parameters}) and parameter inference (see Fig.~\ref{fig:posteriors}) in the investigation of BSM models that serve as possible interpretations of the 2023 PTA signal. We specifically demonstrated that this technique outperforms related but simpler approaches, at least in the case of BSM models that yield a good fit of the NG15 signal. In future work, we are planning to implement the tools introduced here in \texttt{PTArcade}. Besides, it may be interesting to explore similar techniques in the context of CMB physics.  


\medskip\noindent
\textit{Acknowledgments}\,---\,We thank Richard von Eckardstein and Tobias Schr\"oder for collaboration at the early stages of this project and William G.\ Lamb and Rafael R.\ Lino dos Santos for useful discussions. K.\,S.\ is an affililate member of the Kavli Institute for the Physics and Mathematics of the Universe (Kavli IPMU) at the University of Tokyo and as such supported by the World Premier International Research Center Initiative (WPI), MEXT, Japan (Kavli IPMU).
A.\,J.\,I.\ thanks the University of Münster for the kind hospitality during the realization of this project.


\bibliographystyle{JHEP} 
\bibliography{arxiv_1}


\end{document}